# Integrated polarizers based on graphene oxide in waveguides and ring resonators

Jiayang Wu[1], Yunyi Yang[1], Yang Qu[1], Xingyuan Xu[1], Yao Liang[1], Sai T. Chu[2], Brent E. Little[3], Roberto Morandotti[4,5,6], Baohua Jia[1],* and David J. Moss[1]*

[1] Center for Micro-Photonics, Swinburne University of Technology, Hawthorn, VIC 3122, Australia.
E-mail: bjia@ swin.edu.au, dmoss@swin.edu.au.
[2] Department of Physics and Material Science, City University of Hong Kong, Tat Chee Avenue, Hong Kong, China.
[3] Xi'an Institute of Optics and Precision Mechanics Precision Mechanics of CAS, Xi'an, China.
[4] INSR-Énergie, Matériaux et Télécommunications, 1650 Boulevard Lionel-Boulet, Varennes, Québec, J3X 1S2, Canada.
[5] Visiting Professor, ITMO University, St. Petersburg, Russia.
[6] Visiting Professor, Institute of Fundamental and Frontier Sciences, University of Electronic Science and Technology of China, Chengdu 610054, China.

**Abstract**
Integrated waveguide polarizers and polarization-selective micro-ring resonators (MRRs) incorporated with graphene oxide (GO) films are experimentally demonstrated. CMOS-compatible doped silica waveguides and MRRs with both uniformly coated and patterned GO films are fabricated based on a large-area, transfer-free, layer-by-layer GO coating method that yields precise control of the film thickness. Photolithography and lift-off processes are used to achieve photolithographic patterning of GO films with precise control of the placement and coating length. Detailed measurements are performed to characterize the performance of the devices versus GO film thickness and coating length as a function of polarization, wavelength and power. A high polarization dependent loss of ~53.8 dB is achieved for the waveguide coated with 2-mm-long patterned GO films. It is found that intrinsic film material loss anisotropy dominates the performance for less than 20 layers whereas polarization dependent mode overlap dominates for thicker layers. For the MRRs, the GO coating length is reduced to 50 μm, yielding a ~ 8.3-dB polarization extinction ratio between TE and TM resonances. These results offer interesting physical insights and trends of the layered GO films and demonstrate the effectiveness of introducing GO films into photonic integrated devices to realize high-performance polarization selective components.

# 1. Introduction

Polarization control is one of the fundamental requirements in many optical technologies [1-3]. Polarization selective devices, such as polarizers and polarization selective resonant cavities (e.g., gratings and ring resonators), are core components for polarization control in optical systems and find wide applications in polarization-division-multiplexing [4, 5], coherent optical detection [6, 7], photography [8, 9], liquid crystal display [10, 11], and optical sensing [12, 13].

To implement polarization-selective devices, a number of schemes have been proposed and demonstrated, including those based on refractive prisms [14, 15], birefringent crystals [16, 17], fiber components [18-20], and integrated waveguides [21-25]. Among them, integrated polarization-selective devices based on complementary metal-oxide-semiconductor (CMOS) compatible integrated platforms [26] offer advantages of compact footprint, high stability, mass producibility, and high scalability as functional building blocks for photonic integrated circuits (PICs) [2]. Optical waveguides with metal cladding have been widely used to implement waveguide polarizers [27, 28]. Although high polarization-dependent loss (PDL) has been achieved for these polarizers, it usually comes at the expense of high overall propagation loss and requires complicated buffer layers to achieve broadband operation. Recently, the huge optical anisotropy and broadband response of two-dimensional (2D) materials such as graphene and transition metal dichalcogenides have been widely recognized and exploited to implement polarization-selective devices [29-34], including an in-line fiber polarizer with graphene [29], graphene-polymer waveguide polarizers [30, 35], and chalcogenide glass-on-graphene waveguide polarizers [33]. However, none of these demonstrations were based on CMOS compatible platforms. Generally, the integration of 2D materials on CMOS compatible platforms requires layer transfer processes [33, 36], where exfoliated or chemical vapour deposition grown 2D membranes are attached onto dielectric substrates (e.g., silicon and silica wafers). Despite its widespread implementation, the transfer approach itself is complex, which makes it difficult to achieve precise patterning, as well as flexible placement and large-area continuous coating on integrated devices. Accurate control of the layer position, thickness and size is critical for optimizing parameters such as mode overlap and loss for performance. Current methods significantly limit the scale of fabrication for integrated devices incorporating 2D materials.

Owing to its ease of preparation as well as the tunability of its material properties, graphene oxide (GO) has become a highly promising member of the 2D family [37, 38]. Recently [39], a broadband GO-polymer waveguide polarizer with a high PDL of ~ 40 dB was demonstrated, where the GO films were introduced onto an SU8 polymer waveguide using drop-casting methods. The GO film thickness for each drop-casting step was ~ 0.5 µm and the drop coating diameter was ~1.3 mm, neither being ideal for achieving precise control of the placement, thickness, and length of the GO films.

Recently [40, 41], we reported large-area, transfer-free, and high-quality GO film coating on integrated waveguides using a solution-based method with layer-by-layer deposition of GO films. Here, we use these techniques to demonstrate GO-coated integrated waveguide polarizers and polarization-selective micro-ring resonators (MRRs) on a CMOS compatible doped silica platform. We achieve highly precise control of the placement, thickness, and length of the GO films coated on integrated photonic devices by using our layer-by-layer GO coating method followed by photolithography and lift-off processes. The latter overcomes the layer transfer fabrication limitations of 2D materials and represent a significant advance towards manufacturing integrated photonic devices incorporated with 2D materials. We measure the performance of the waveguide polarizer for different GO film thicknesses and lengths versus polarization, wavelength, and power, achieving a very high PDL of ~ 53.8 dB. For GO-coated integrated MRRs, we achieve an 8.3-dB polarization extinction ratio between the TE and TM resonances, with the extracted propagation loss showing good agreement with the waveguide results. Furthermore, we present layer-by-layer characterization of the linear optical properties of 2D layered GO films, including detailed measurements that conclusively determine the material loss anisotropy of the GO films as well as the relative contribution of film loss anisotropy versus polarization-dependent mode overlap, to the device performance. These results offer interesting physical insights and trends of the layered GO films from monolayer to quasi bulk like behavior and confirm the high-performance of integrated polarization selective devices incorporated with GO films.

## 2. GO-coated Waveguide Polarizer

### 2.1 Device Fabrication

**Figure 1(a)** shows a schematic of a uniformly GO-coated waveguide polarizer. The waveguides were fabricated from high-index doped silica glass core surrounded by silica via CMOS compatible processes [26, 42] with chemical mechanical polishing (CMP) used as the last step to remove the upper cladding, so as to enable GO film coating on the top surface of the waveguide. GO coating was achieved with a solution-based method that yielded layer-by-layer GO film deposition, as reported previously [40, 41]. Four steps for in-situ assembly of monolayer GO films were repeated to construct multilayer GO films on the desired substrate, with the process being highly scalable.

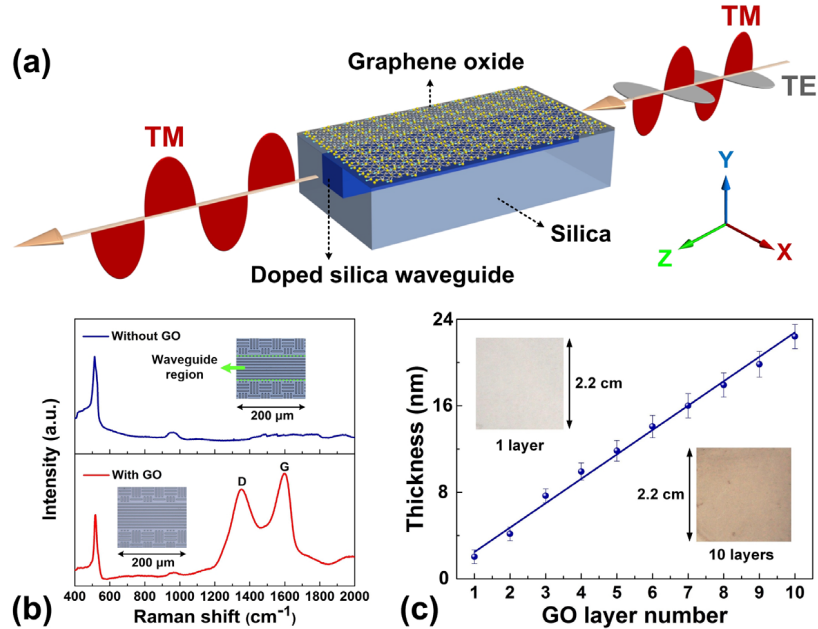

**Figure 1.** (a) Schematic illustration of GO-coated integrated waveguide polarizer. (b) Raman spectra of the integrated chip without GO and with 2 layers of GO. Insets show the corresponding microscope images with 9 parallel waveguides in the guiding region. (c) Measured GO film thickness versus GO layer number. Insets show the images of a 2.2 cm × 2.2 cm silica substrate coated with 1 and 10 layers of GO, respectively.

We uniformly coated waveguides with 1 to 10 layers of GO. **Figure 1(b)** shows the measured Raman spectra of the waveguides without GO and with 2 layers of GO, confirming the integration of GO onto the top surface by the presence of the D (1345 cm$^{-1}$) and G (1590 cm$^{-1}$) peaks of GO. The microscope images of the integrated waveguide with zero and 2 layers of GO are shown in the insets, which illustrate the good morphology of the GO films. **Figure 1(c)** shows the thickness of GO films versus the layer number characterized by atomic force microscopy. The insets show images of 1 and 10 layers of GO coated on a 2.2 cm × 2.2 cm silica substrate with high uniformity. The dependence of GO film thickness versus layer number shows a nearly linear relationship, with a thickness of ~2.18 nm on average for each layer.

In addition to the uniformly coated devices, we selectively patterned areas of GO films using lithography and lift-off processes. Apart from allowing precise control of the size and placement of the GO films, this enabled us to test the performance of the GO-coated waveguide polarizers with shorter GO coating lengths but higher film thicknesses (up to 100 layers). The chip was first spin-coated with photoresist and then patterned using photolithography to open a window on the waveguide. Next, GO films were coated on the chip using the method mentioned above and patterned via a lift-off process.

As compared with the drop-casting method that has a GO film thickness of ~ 0.5 μm and a minimum size of about 1.3 mm for each step [39], the combination of our GO coating method with photolithography and lift-off allows precise control of the film placement, size, and thickness (with an ultrahigh resolution of ~2.18 nm), all of which are critical for optimizing the device performance including the polarization figure of merit (FOM) and four-wave mixing conversion efficiency. Further, our solution based GO coating approach, unlike for example, the sophisticated transfer processes employed for coating 2D materials such as graphene [29, 43], is capable of covering large areas (e.g., a 4 inch wafer) on dielectric substrates (e.g., silicon and silica wafers) with relatively few defects. The combination of patterning and deposition control of GO films along with large area coating capability is critical for large-scale integrated devices incorporated with GO.

*2.2 Polarization Loss Measurements*

We used an 8-channel single-mode fiber (SMF) array to butt couple both TE and TM polarized continuous-wave (CW) light from a tunable laser near 1550 nm into the waveguides. The mode coupling loss between the SMF array and the waveguides was ~8 dB/facet, which can readily be reduced to ~1.0 dB/facet with mode convertors [26]. The propagation loss of the uncoated 1.5-cm-long waveguides was very low (< 0.25 dB/cm) and so the total insertion loss (TE= −16.2 dB; TM = −16.5 dB) of the uncoated devices was dominated by mode coupling loss. The slight PDL of the uncoated waveguides resulted mainly from a slightly different mode-coupling mismatch but possibly also polarization dependent scattering loss from the roughness of the polished top surface.

To characterize the performance of the devices, we introduce two figures of merit (FOMs) – one for the excess propagation loss ($FOM_{EPL}$) and one for the overall polarization dependent loss ($FOM_{PDL}$):

$$FOM_{EPL} = (EPL_{TE} - EPL_{TM}) / EPL_{TM}, \qquad (1)$$
$$FOM_{PDL} = PDL / EIL, \qquad (2)$$

where the excess propagation losses, $EPL_{TE}$ (dB/cm) and $EPL_{TM}$ (dB/cm), are GO-induced excess propagation losses for the TE and TM polarizations, respectively. PDL (dB) is the polarization dependent loss defined as the ratio of the maximum to minimum insertion losses. The excess insertion loss, EIL (dB), is the insertion loss induced by the GO film over the uncoated waveguide. The EIL only considers the insertion loss induced by GO, while excluding from the overall insertion loss both the mode coupling loss between the SMF array and the waveguide as well as the propagation loss of the uncoated waveguide. In our case, since the TM polarization had the lowest insertion loss, EIL is the excess GO-induced insertion loss for the TM polarization and is given by EIL= $EPL_{TM}$ · L, where L is the GO coating length. Note that $FOM_{EPL}$ only considers the propagation loss difference induced by the GO films, and so is more accurate for the characterization of their material anisotropy, whereas $FOM_{PDL}$ is more widely used for evaluating the device performance [33] since it also includes the background (uncoated) PDL. $FOM_{EPL}$ equals $FOM_{PDL}$ only when the TE and TM polarized insertion losses of the uncoated waveguide are the same.

**Figure 2** shows the polarization dependent (TE (in-plane) and TM (out-of-plane)) performance for both the 1.5-cm-long uniformly coated waveguides (0-10 layers), left side (i), and the patterned 2-mm-long devices (10-100 layers), right side (ii). **Figure 2(a)** shows the polarization dependent insertion loss. The data points depict the average values obtained from the experimental results of 3 duplicate devices and the error bars illustrate the variations for different samples. **Figure 2(b)** presents the PDL and EIL calculated based on the average insertion loss in **Figure 2(a)**. **Figures 2(c)** and **(d)** show the FOMs calculated based on **Figure 2(b)** and the polar diagrams for the insertion loss, respectively.

The TE insertion loss increases much more strongly than TM with layer number, thus yielding a large PDL with low EIL and forming the basis for our high-performance polarization dependent devices. Since GO is a dielectric film, our TM-pass GO-coated waveguide polarizer is quite different from TE-pass metal-clad waveguide polarizers based on a deeper power penetration of the evanescent TM field into a lossy metal cladding [27, 28]. The PDL reached a maximum of ~37.4 dB for a 10 layer uniformly coated device and ~53.8 dB for a 100 layer patterned device, with a modest maximum EIL of ~5.0 dB and ~7.5 dB for the two devices, respectively. By optimizing the waveguide geometry to achieve a better mode overlap with the GO films [33], the EIL can be further reduced. Moreover, the PDL was still increasing at a rate of 2-3 dB/cm/layer at 100 layers, and so substantially higher PDL can be obtained using layers thicker than 200 nm. Both $FOM_{EPL}$ and $FOM_{PDL}$ increase with a maximum of $FOM_{EPL}$ and $FOM_{PDL}$ of ~8.2 and ~8.1, respectively, at about 50 layers with the difference between them subsequently decreasing. This is because the impact of the background PDL (~0.3 dB) becomes smaller as the EIL increases for increased GO layer numbers.

*2.3 GO Film Properties*

**Figure 3(a)** shows the experimental propagation loss (dB/cm) extracted from Fig. 2(a) for both polarizations of the two devices, along with the TE propagation loss calculated (by means of the Lumerical FDTD commercial mode solving software) using ellipsometry measurements (at 1550 nm) for the refractive index *n* and extinction coefficient *k* of two samples having 2 layers (**Figure 3(a-i)**) and 20 layers (**Figure 3(a-ii)**) of GO. Since the out-of-plane (TM polarized) response of the layered GO films is much weaker [44, 45], we could only measure, via ellipsometry, the in-plane (TE polarized) *n* and *k* of the GO films (uncertainty < 3%), which were used (in conjunction with the mode solving software) to calculate the waveguide loss for the TE polarization. The simulations assume constant *n*, *k* for different GO layer numbers in each plot. In **Figure 3(a)**, the experimental TE propagation loss agrees extremely well with simulations for 2 and 20 layers of GO. For other thicknesses the experimental TE loss increased more rapidly with GO layer number, indicating that the intrinsic GO film loss increases with thickness. This is not surprising, and could be due to any number of effects such as increased scattering loss and absorption induced by imperfect contact between the multiple GO layers as well as interactions between the GO layers.

**Figure 3(b)** shows the propagation loss per layer for both devices given by the loss divided by the number of layers. The TE loss starts at 1 dB/cm/layer at 1−2 layers, rising to 2.5−3 dB beyond 10 layers. The loss per layer agrees well for both devices, in the case of 10 layers. The TE loss per layer increases up to about 50 layers, after which it levels off and decreases marginally at the highest number of layers. This could be due to any number of factors including larger scattering loss and absorption among the multiple GO layers with increasing layer thickness. TM polarization, on the other hand, has a much lower film loss of 0.1−0.4 dB/cm/layer for both samples over all thicknesses.

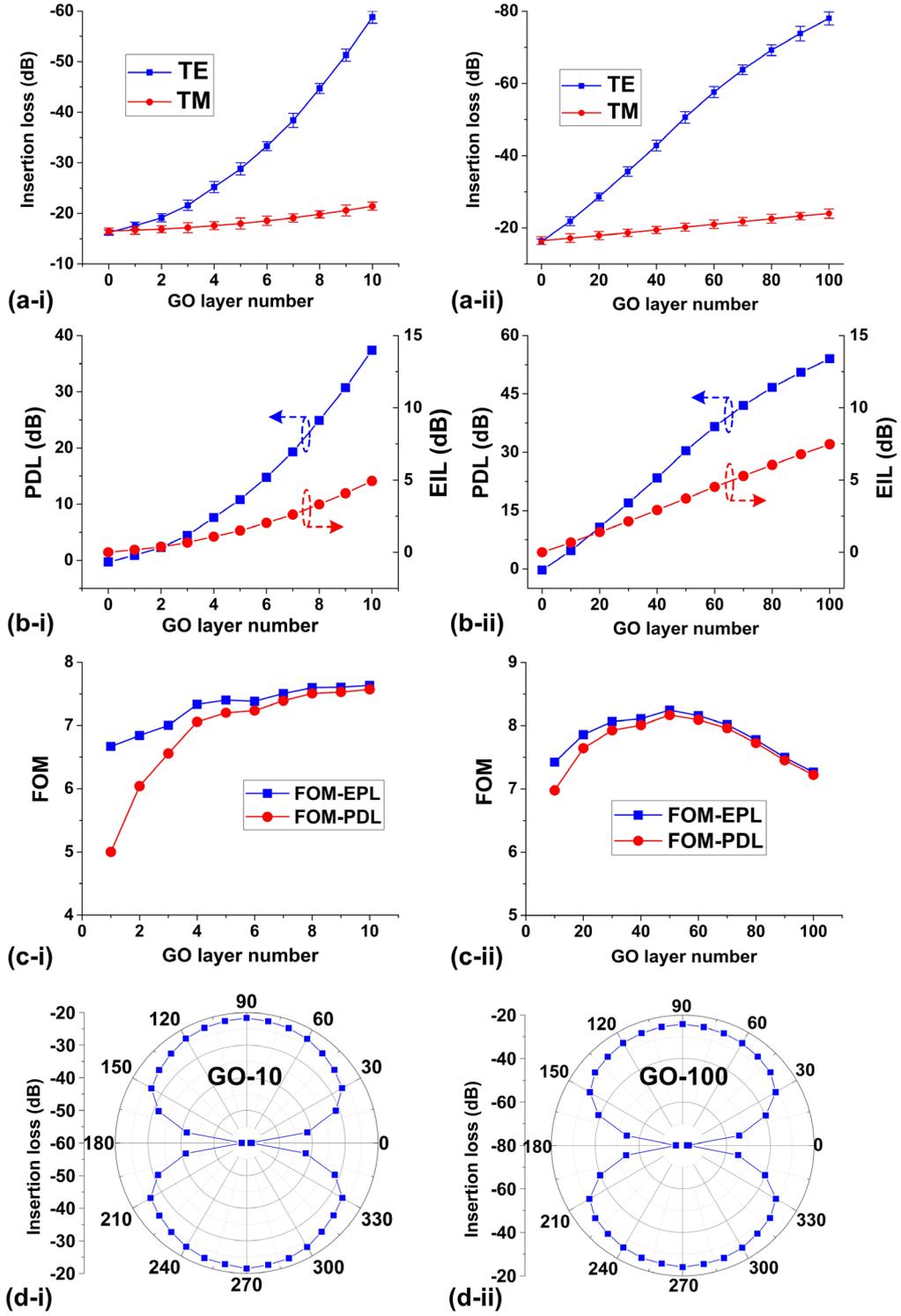

**Figure 2.** (a) Measured TE and TM polarized insertion loss. (b) Extracted polarization dependent loss (PDL) and excess insertion loss (EIL). (c) Calculated figure of merits (FOMs). In (a) − (c), (i) shows the results for 1.5-cm-long uniformly coated waveguides (0, 1, 2, ..., 10 layers of GO) and (ii) shows the results for 1.5-cm-long waveguides with 2-mm-long patterned GO (0, 10, 20, ..., 100 layers of GO). (d) Polar diagrams presenting the polarizer performance for (i) 1.5-cm GO coating length, 10 layers of GO and (ii) 2-mm GO coating length, 100 layers of GO. The polar angle represents the angle between the input polarization plane and the substrate. The input CW power and wavelength in (a) and (d) are 0 dBm and 1550 nm, respectively.

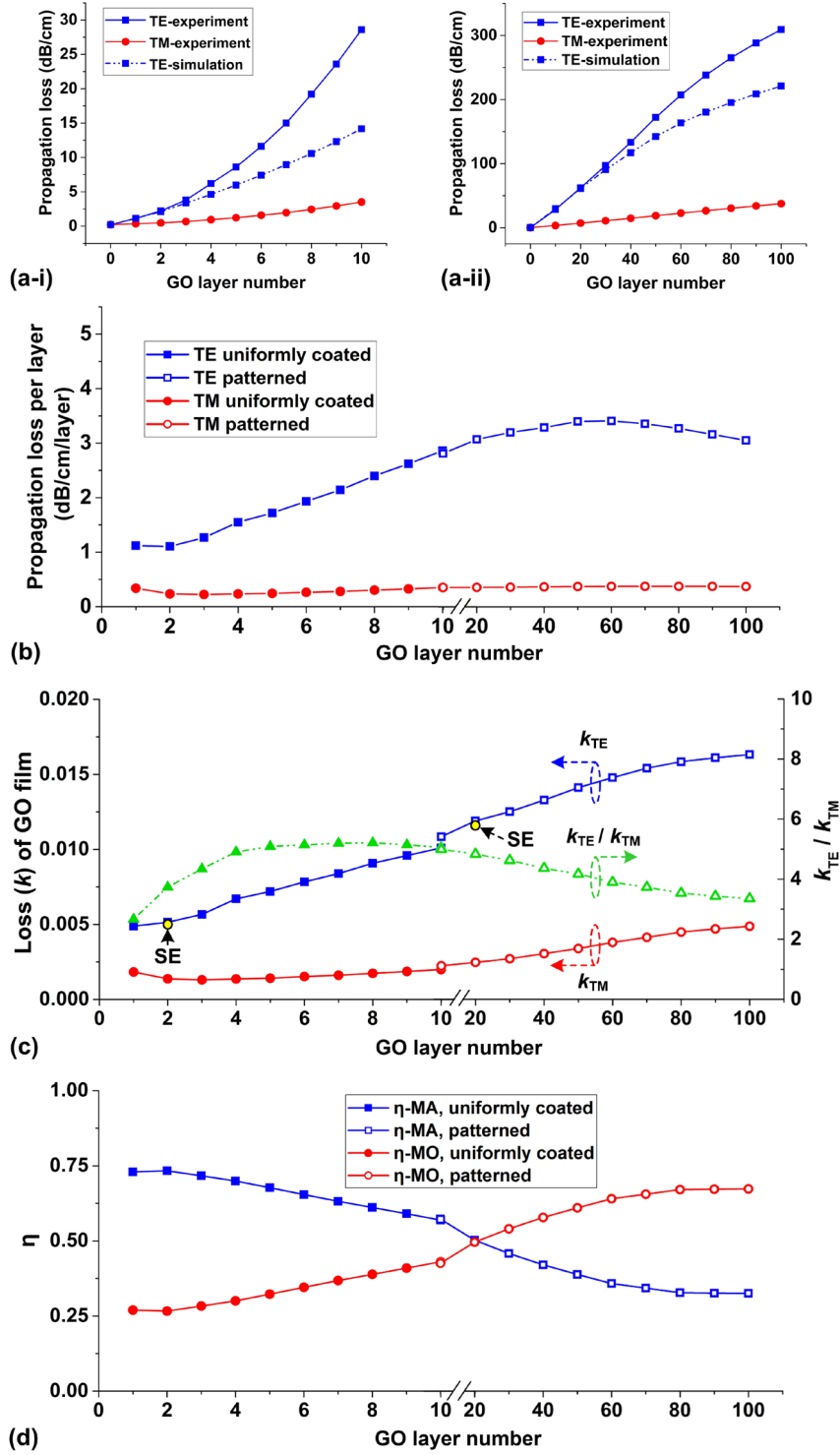

**Figure 3.** (a) Experimental waveguide propagation loss (extracted from Fig. 2(a)) for (i) a uniformly coated device with 0−10 layers of GO and (ii) a patterned device with 10−100 layers of GO. Simulated TE loss estimated by using in-plane (TE polarized) $n$ and $k$ for (i) 2 and (ii) 20 layers of GO obtained from ellipsometry measurements in conjunction with mode solving software (assuming constant $n$, $k$ for different layer numbers in each plot). (b) Experimental loss per layer given by the experimental loss in (a) divided by the number of layers for both polarizations. (c) Material loss of the GO films for TE ($k_{TE}$) and TM ($k_{TM}$) polarizations as well as their ratio ($k_{TE}$ / $k_{TM}$). The two yellow data points (labelled as "SE") show the in-plane $k$ measured by spectral ellipsometry (SE) for 2 and 20 layers of GO. (d) Fractional contribution to PDL from mode overlap ($\eta_{MO}$) and material loss anisotropy ($\eta_{MA}$). In (b) − (d), the solid data points refer to the results for a uniformly coated device, whereas the hollow points are for a patterned device.

**Figure 3(c)** shows the GO film loss ($k$) extracted from the waveguide loss measurements. We neglect the effect of any variations in GO film refractive index ($n$) on the mode overlap. The much larger film loss $k$ for TE polarization ($k_{TE}$) is a reflection of the high intrinsic loss anisotropy of the GO films, which has been previously noted in a qualitative manner [46]. The results presented here are the first quantitative measurement of intrinsic film loss anisotropy in GO films, showing that the in-plane (TE) electric field loss is significantly higher than the out-of-plane (TM) loss. This material anisotropy significantly improves the polarization dependent loss of our devices, which otherwise would derive solely from the polarization dependent mode overlap with the GO films. Both $k_{TE}$ and $k_{TM}$ increase with layer number, with the ratio $k_{TE}$ / $k_{TM}$ reaching a maximum of about 5 at 4−10 layers after which it decreases to ~3 for very thick films (100 layers). This could arise from a reduction in, for example, anisotropic film stress related effects for very thick films (>200 nm). The two $k$ data points (marked in yellow) in **Figure 3(c)** are obtained from ellipsometry measurements, with $k_{ellips}^{2\text{-layers}} = 0.0050$ and $k_{ellips}^{20\text{-layers}} = 0.01165$, thus agreeing extremely well with the waveguide propagation loss results − $k_{wg}^{2\text{-layers}} = 0.0054$ and $k_{wg}^{20\text{-layers}} = 0.01212$. Note also that the $k$'s for the patterned and uniformly coated waveguides agree well for the same number (10) of GO layers.

**Figure 3(d)** shows the fractional contribution to the PDL from the polarization dependent mode overlap ($\eta_{MO}$) and the material film loss anisotropy ($\eta_{MA}$), where $\eta_{MO} + \eta_{MA} = 1$. We calculated $\eta_{MO}$ and $\eta_{MA}$ by comparing the measured propagation loss in **Figure 3(a)** for both polarizations, to the propagation loss calculated assuming isotropic film properties (using the extracted $k_{TE}$ in **Figure 3(c)** for both TE and TM polarizations). For low GO layer numbers (<10) the PDL is dominated by the material anisotropy $\eta_{MA}$ at 75%, despite the overall TE loss only being 1 dB/cm/layer (at 1−2 layers). The contribution of the GO material anisotropy steadily decreases, becoming comparable to the mode overlap contribution, $\eta_{MO}$, at ~ 20 layers, beyond which, for very thick films (100 layers), is smaller than $\eta_{MO}$, which is about 65%. This could reflect the transition of the film properties slightly towards a bulk (isotropic) material for very thick films. However, it is also interesting to note that even for very thick films the intrinsic film loss anisotropy is still large enough to form the basis for polarization dependent devices. To reconfigure the polarization selection, it would be relatively easy to change the mode overlap, while changing the material loss anisotropy is more challenging. A possible method to implement a TE-pass GO waveguide polarizer would be to conformally coat a high-aspect-ratio waveguide with GO films on the sidewalls. Finally, we note that although our GO-coated waveguide polarizers were based on a CMOS compatible doped silica platform, these GO films can readily be introduced into other integrated platforms (e.g., silicon and silicon nitride) [47-49], offering polarization selective devices with reduced footprint.

*2.4 Optical Bandwidth and Power Dependence*

**Figures 4(a-i)** and **(a-ii)** illustrate the PDL of both uniformly coated and patterned waveguides versus wavelength from 1500 to 1600 nm, showing a variation less than 2 dB and confirming the broadband operation of the polarizers. Table I compares the performance of a range of silicon photonic polarizers and 2D material based optical polarizers. The bandwidth of the material anisotropy of GO thin films is very broad [37] − several hundred nanometers, even extending to visible wavelengths (**Figure 4(a-i)** insets). This is a distinct advantage of GO-coated waveguide polarizers that is extremely challenging to achieve with silicon photonic polarizers [3, 21, 24]. Moreover, GO based polarizers have simpler designs with higher fabrication tolerance as compared with silicon photonic polarizers. Indeed, the latter require precise design and control of the dimensions [2, 3]. It is also interesting to note that the PDL slightly increases at longer wavelengths for both the uniformly coated and patterned devices. This is probably a result of the excitation of high-order modes in the doped silica waveguides at shorter wavelengths, which reduces the strength of the interaction between the GO films and the evanescent field leaking from the waveguides, thus leading to a degradation in PDL.

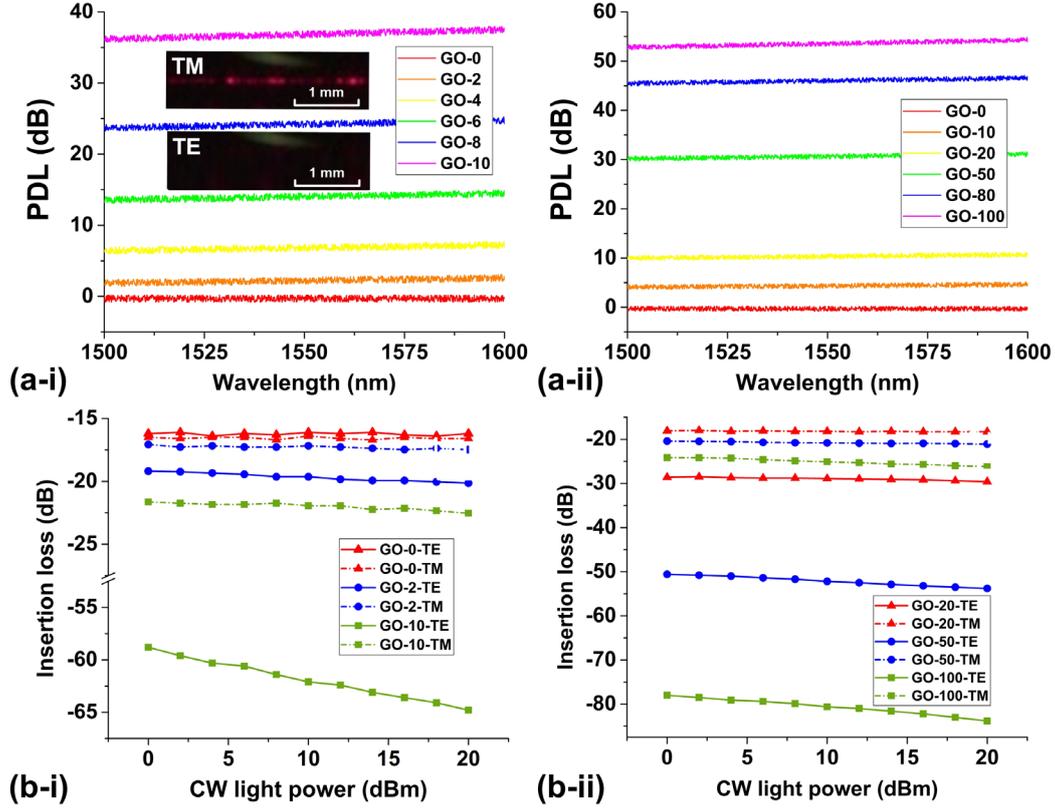

**Figure 4.** Measured PDL in the wavelength range of 1500 nm ~1600 nm for (i) 1.5-cm GO coating length, 0−10 layers of GO and (ii) 2-mm GO coating length, 0−100 layers of GO. The input CW power is 0 dBm. Insets in (i) show optical images of TM and TE polarized light at a visible wavelength of ~632 nm along the waveguide uniformly coated with 10 layers of GO. (c) Measured TE and TM polarized insertion losses at different input CW powers for (i) 1.5-cm GO coating length, 0−10 layers of GO and (ii) 2-mm GO coating length, 0−100 layers of GO. The input CW wavelength is 1550 nm.

The insertion loss versus CW power is shown in **Figures 4(b-i)** and **(b-ii)**, for both samples and polarizations, indicating only a slight increase in loss for the thicker layers and only for the TE polarization, possibly due to photo-thermal reduction of the GO film at higher powers [50]. This might also result from self-heating and thermal dissipation in the multilayer GO film, being the subject of on-going research. In contrast, since the TM polarized absorption was very low, it did not show any significant variation with power. The increase in loss for TE polarized light was reversible − indicating that the optically induced changes were reversible. Note that this slight reversible increase in loss for the TE polarization with power actually enhances the device performance. Finally, we have shown previously [38, 51] that the material properties of GO can also be permanently changed by laser-induced photo-reduction but at significantly higher power levels than those used here, with femtosecond laser pulses. This is different from the reversible photo-thermal reduction observed here.

# Table I.

**Performance comparison of silicon photonic polarizers and 2D material based polarizers. PDL: polarization dependent loss. IL: insertion loss. FOM: figure of merit. ER: extinction ratio. N/A: not applicable.**

| Device | Length [mm] | PDL [dB] | IL[a] [dB] | FOM =PDL/IL | Bandwidth [nm] (≥ 20 dB ER) |
|---|---|---|---|---|---|
| Shallowly etched silicon waveguide [21] | 1 | 25 | 3 | 8.3 | >100 |
| Silicon subwavelength grating waveguide [24] | 0.009 | 27 | 0.5 | 54 | ~60 |
| Silicon nanoplasmonic slot waveguide [22] | 0.001 | 16 | 2.2 | 7.3 | ER<20 |
| Graphene-coated side polished fiber [29] | 2.1 | 27 | 5 | 5.4 | ~1000 |
| Graphene-coated polymer waveguide [30] | 7 | 19 | 26 | 0.73 | N/A[b] |
| Chalcogenide glass-on-graphene waveguide [33] | 0.4 | 23 | 0.8 | 29 | ~450 |
| $MoS_2$-coated Nd:YAG waveguide [52] | 10 | 3 | 0.4 | 7.5 | ER<20 |
| GO-coated polymer waveguide [39] | 1.3 | 40 | 6.5 | 6.1 | >490 |
| GO-coated doped silica waveguide (this work) | 2 | 54 | 7.5 | 7.2 | >540[c] |

a) The ILs exclude the fiber-to-chip coupling losses.
b) The device was only characterized at a single wavelength.
c) We cannot precisely characterize the bandwidth due to the lack of suitable lasers. In our measurements, we achieve high PDLs of over 52.4 dB from 1500 to 1600 nm (with a variation less than 2 dB) and also a PDL of ~25.2 dB at 1064 nm.

## 3. Polarization-Selective Microring Resonators

We coated GO films onto integrated MRRs to implement polarization-selective MRRs, for applications such as polarization-handling devices in coherent receivers [53]. **Figure 5(a)** shows a schematic of the GO-coated polarization selective MRR, with the insets showing schematic atomic structure of GO and scanning electron microscope (SEM) image of the GO film with up to 5 layers of GO. The unclad MRR made from high-index doped silica glass was fabricated via the same CMOS compatible processes as for the integrated waveguides in **Section 2** [26, 54]. The ring and the bus waveguide had the same waveguide geometry as in **Section 2**. The radius of the MRR was ~592 μm, corresponding to a free spectral range of ~0.4 nm (~50 GHz). The gap between the ring and the bus waveguide was ~0.8 μm. We used the same method to couple CW light to the MRR. We fabricated and tested two types of GO-coated MRR polarizers, uniformly coated with 1−5 layers of GO and patterned (50 μm in length) with 10−100 layers of GO using the same photolithography and lift-off processes as for the patterned waveguide in **Section 2**. Gold markers, prepared by metal lift-off after photolithography and electron beam evaporation, were used for precise alignment and accurate placement of GO on the MRR (deviation < 20 nm). Microscope images of the integrated MRR uniformly coated with 5 layers of GO and patterned with 50 layers of GO are presented in **Figures 5(b)** and **(c)**, respectively. Note that although a number of concentric rings are shown, only the center ring was coupled with the through/drop bus waveguides to form a MRR − the rest were simply used to enable easy identification by eye. There are several factors that can limit the minimum pattern length, such as the thickness of the GO film, lithography resolution, size of the GO flakes, and thickness of the photoresist. By using oxidation and vigorous ultrasonics [41], we achieved ultrasmall GO flake sizes down to ~50 nm. For thin GO films (< 10 layers), the pattern length was mainly limited by the lithography resolution and GO flake size, whereas for thick GO films (> 50 layers), the thickness itself becomes the dominant factor. By using e-beam lithography to write patterns on a 300-nm-thick photoresist, we achieved short pattern lengths of ~150 nm and ~500 nm for 2 layers and 30 layers of GO, respectively. This confirms the high quality and resolution of the GO deposition and patterning process as well as the good adhesion between the GO film and the integrated devices.

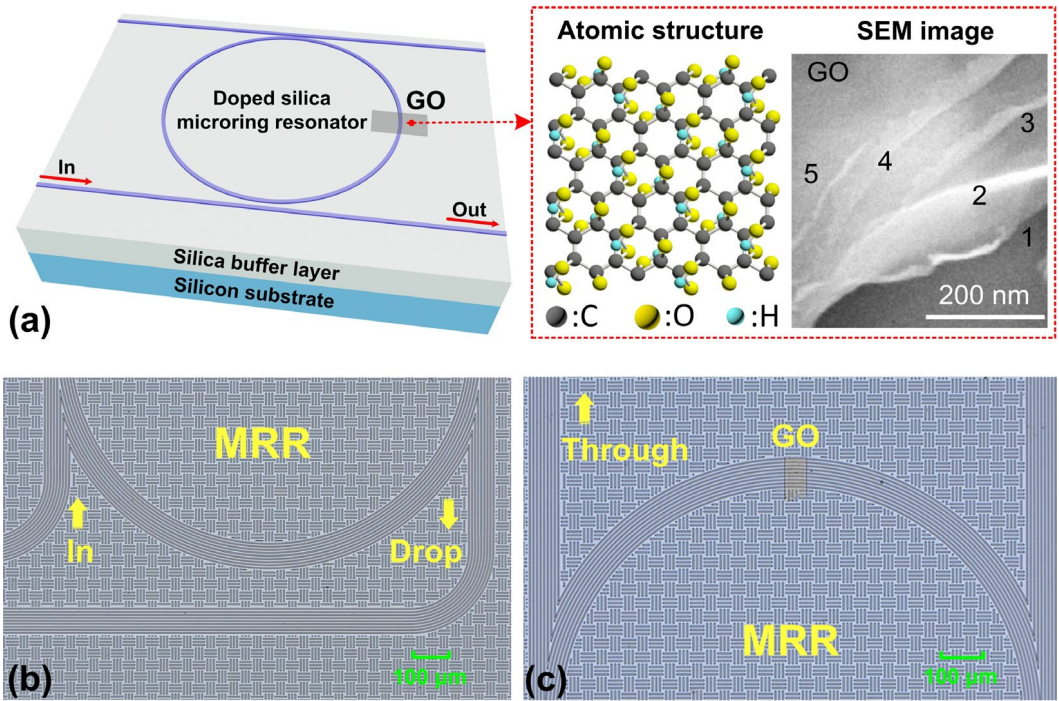

**Figure 5.** (a) Illustration of GO-based polarization-selective MRR. Insets show schematic atomic structure of GO and a scanning electron microscope (SEM) image of layered GO film. The numbers in the SEM refer to the number of layers for that part of the image. (b)−(c) Microscope images of the integrated MRR uniformly coated with 5 layers of GO and patterned with 50 layers of GO, respectively.

The measured TE and TM polarized transmission spectra of the uniformly GO-coated MRR are shown in **Figures 6(a)** and **(b)**, respectively, while the transmission spectra of the patterned MRR are shown in **Figures 7(a)** and **(b)**, all measured with the same doped silica MRR at a CW power of ~0 dBm. The resulting Q factors and extinction ratios are shown in **Figure 8(a)**. The uncoated MRR had high extinction ratios (> 15 dB) and relatively high Q factors (180,000) (although significantly less than for buried waveguides [42]) for both polarizations. Those values decreased with GO layer number – particularly for the TE polarization, as expected. For the patterned MRR with 50 layers of GO, a maximum polarization extinction ratio (defined as the difference between the extinction ratios of the TE and TM polarized resonances) of ~8.3 dB was achieved. This can be further improved by optimizing the waveguide geometry, GO film thickness, and coating length to properly balance the mode overlap and material anisotropy. The propagation loss of the GO hybrid integrated waveguides for TE and TM polarizations was obtained using the scattering matrix method to fit the measured spectra in **Figures 6** and **7** and is shown for uniformly coated (0−5 layers) and patterned rings (0−100 layers) in **Figures 8(b-i)** and **(b-ii)**, respectively, along with the waveguide propagation loss obtained from the waveguide experiment (i.e., the experimental propagation loss in **Figure 3(a)**). Since different resonances did not show a significant variation over small wavelength ranges, we only fit one resonance around 1549.5 nm in each measured spectrum. The fit coupling coefficients between the ring and the bus waveguide for TE and TM polarizations are ~0.241 and ~0.230, respectively. The close agreement reflects the stability and reproducibility of our layer-by-layer GO film coating method. We also note that the propagation loss obtained from the ring resonator experiment is slightly higher than that obtained from the waveguide experiment, especially for the TE polarization. This probably results from photo-thermal reduction of GO in the resonant cavity at higher intensity.

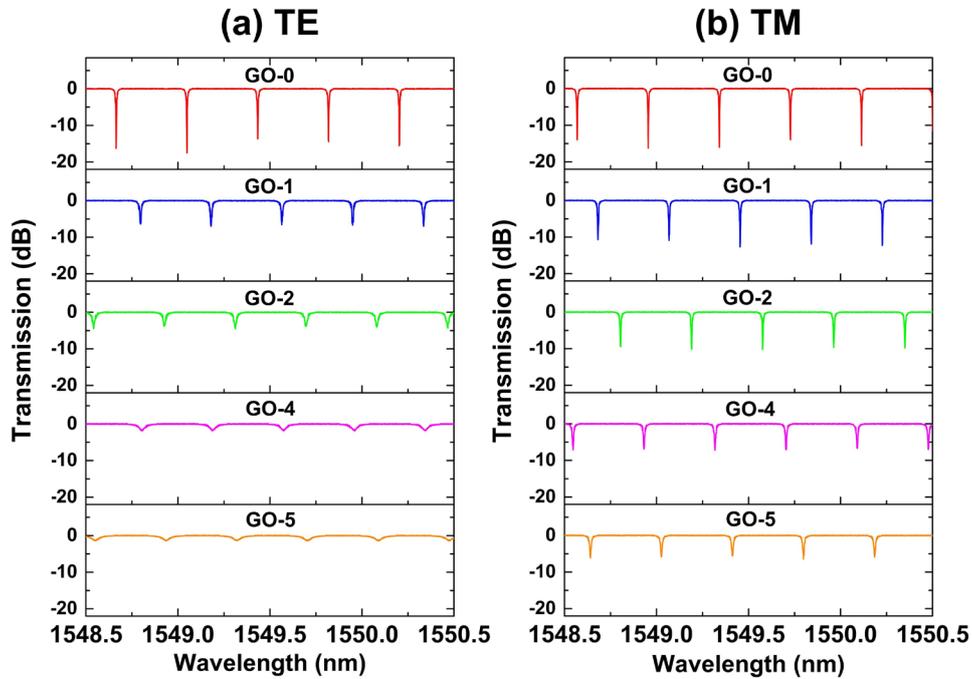

**Figure 6.** Measured (a) TE and (b) TM polarized transmission spectra of the MRR uniformly coated with 0−5 layers of GO.

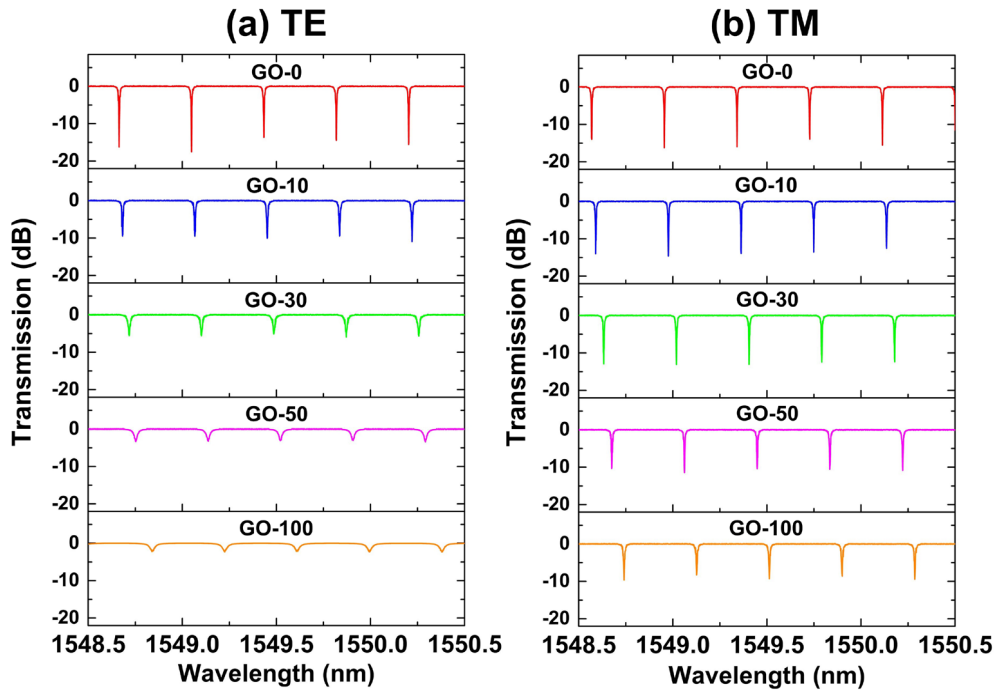

**Figure 7.** Measured (a) TE and (b) TM polarized transmission spectra of the patterned MRR with 0−100 layers of GO.

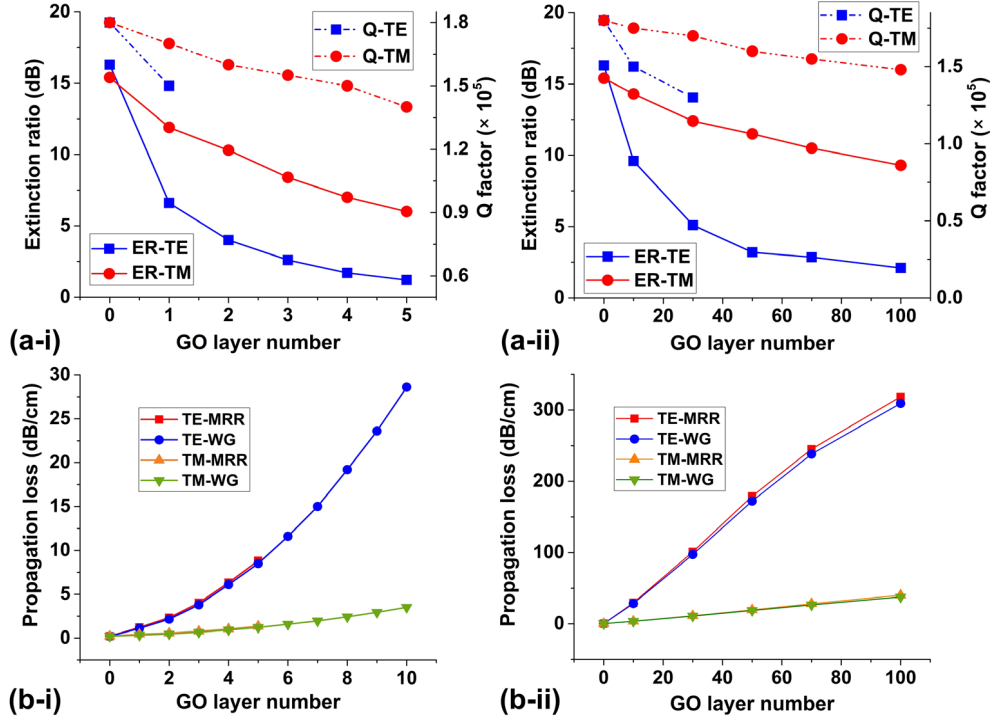

**Figure 8.** (a) Fit extinction ratios (ERs) and Q factors (Q) for the MRR (i) uniformly coated with 0−5 layers of GO and (ii) patterned with 0−100 layers of GO. The Q factors are not shown when the ER is < 5 dB. (b) Fit propagation loss obtained from the MRR experiment and extracted propagation loss obtained from the waveguide (WG) experiment. (i) 0−10 layers of GO, (ii) 0−100 layers of GO.

## 4. Conclusion

We demonstrate waveguide polarizers and polarization-selective MRRs incorporated with layered GO films. We achieve precise control of the placement, thickness, and length of the GO films using layer-by-layer coating of GO films followed by photolithography and lift-off. We achieve a high PDL of ~53.8 dB for the patterned GO-coated waveguide, and for the GO-coated integrated MRR an ~8.3-dB polarization extinction ratio between the TE and TM resonances. We find that the PDL is dominated by material loss anisotropy of the GO film for thin films, and by polarization dependent mode overlap for thick films. These integrated GO hybrid waveguide polarizers and polarization-selective MRRs offer a powerful new way to implement high-performance polarization selective devices for large-scale PICs.

## Acknowledgements

J. Wu and Y. Yang contribute equally to this paper. This work was supported by the Australian Research Council Discovery Projects Programs (No. DP150102972 and DP150104327) and the Swinburne ECR-SUPRA program. We also acknowledge the Swinburne Nano Lab for the support in device fabrication and characterization. RM acknowledges support by the Natural Sciences and Engineering Research Council of Canada (NSERC) through the Strategic, Discovery and Acceleration Grants Schemes, by the MESI PSR-SIIRI Initiative in Quebec, and by the Canada Research Chair Program. He also acknowledges additional support by the Government of the Russian Federation through the ITMO Fellowship and Professorship Program (grant 074-U 01) and by the 1000 Talents Sichuan Program in China.